\documentclass{article}
\usepackage{spconf,amsmath,graphicx,hyperref}
\usepackage{multirow}
\usepackage{subcaption}
\usepackage{latexsym}
\usepackage{microtype}
\usepackage{inconsolata}
\usepackage{algpseudocode}
\usepackage{colortbl}
\usepackage{siunitx}
\usepackage{booktabs}
\usepackage{makecell}
\usepackage[numbers, sort&compress]{natbib}

\usepackage{xurl} 

\title{Nord-Parl-TTS: Finnish and Swedish TTS Dataset \\ from Parliament Speech}
\name{Zirui Li$^{\star}$ \qquad Jens Edlund$^{\dagger}$ \qquad Yicheng Gu$^{\ddagger}$ \qquad Nhan Phan$^{\star}$ \qquad Lauri Juvela$^{\star}$ \qquad Mikko Kurimo$^{\star}$}
  
  \address{$^{\star}$ Department of Information and Communications Engineering, Aalto University, Espoo, Finland \\
      $^{\dagger}$Speech, Music \& Hearing, KTH Royal Institute of Technology, Stockholm, Sweden \\
      $^{\ddagger}$School of Data Science, The Chinese University of Hong Kong, Shenzhen, China}
\begin{document}
%
\maketitle
\begin{abstract}

Text-to-speech (TTS) development is limited by scarcity of high-quality, publicly available speech data for most languages outside a few high-resource languages.  
We present \textit{Nord-Parl-TTS}, an open TTS dataset for Finnish and Swedish based on speech found in the wild.
Using recordings of Nordic parliamentary proceedings, we extract 900 hours of Finnish and 5090 hours of Swedish speech suitable for TTS training.
The dataset is built using an adapted version of the Emilia data processing pipeline and includes unified evaluation sets to support model development and benchmarking.
By offering open, large-scale data for Finnish and Swedish, Nord-Parl-TTS narrows the resource gap in TTS between high- and lower-resourced languages.

\end{abstract}

\section{Introduction}
\label{sec:intro}

\begin{figure*}[t]
  \centering
  \includegraphics[width=.9\linewidth]{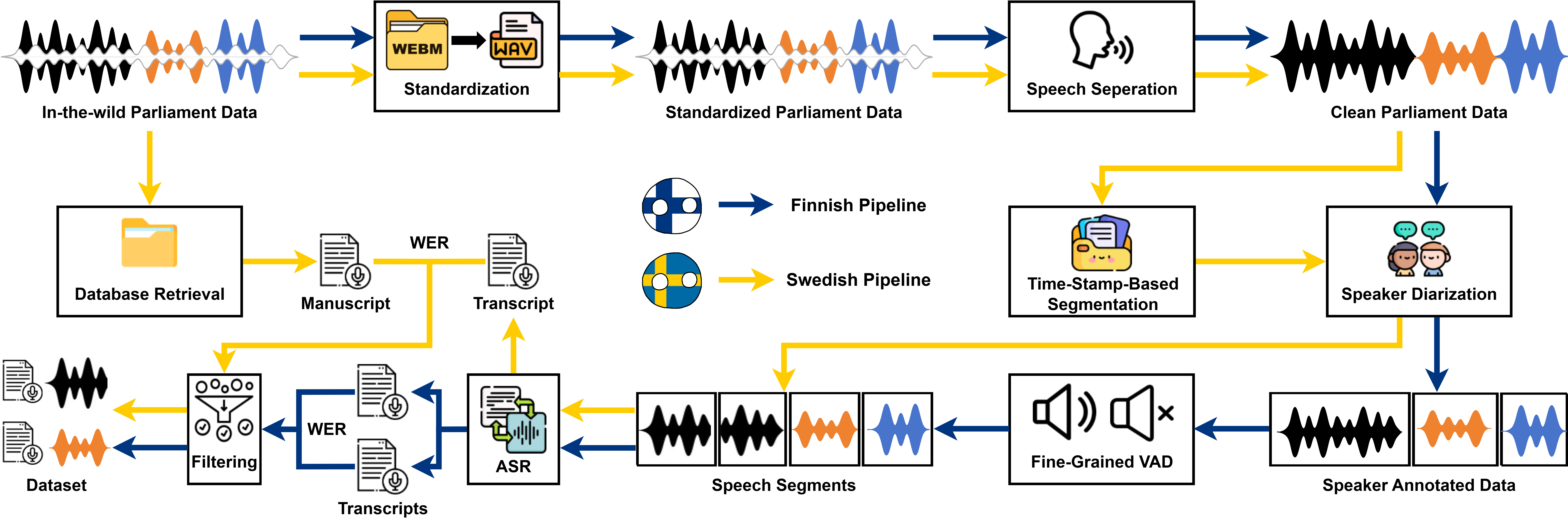}
  \caption{An overview of the proposed \textit{Nord-Parl-TTS} data processing pipeline.
  }
  \label{fig:pipeline}
\end{figure*}



Text-to-speech (TTS) converts written text into speech. With current methods, the humanlikeness and intelligibility of the resulting speech depend on the availability of speech datasets.
When trained on extensive, high-quality data, modern TTS systems can generate speech with intelligibility and prosodic variation comparable to human speakers in high-resource languages~\cite{chen2024f5, wang2025maskgct}.
Conventional TTS datasets rely on studio recordings, typically drawn from audiobooks or produced by native speakers and voice actors~\cite{ljspeech17, zen2019libritts, Yamagishi2019CSTRVC}.
Low-resource languages lack the publicly available recordings and speaker access required to build conventional high-quality TTS datasets. As a result, system development is slow and costly.

Recent work on building extensive and diverse datasets from speech found in the wild~\cite{he2024emilia, ma2024wenetspeech4tts, liu2025voxpopulitts, jung25c_interspeech} directly targets this challenge. In particular, Emilia~\cite{he2024emilia} proposed a unified data processing pipeline and released 101k hours of multilingual speech data. This was followed by several works~\cite{ma2024wenetspeech4tts, liu2025voxpopulitts, jung25c_interspeech} that further extended the data quality and quantity. While these efforts mainly target high-resource languages like English and Mandarin Chinese, languages with lower resources are often overlooked. 
As a step toward addressing this imbalance in the Nordic region, we present \textit{Nord-Parl-TTS}, an open in-the-wild TTS dataset for Finnish and Swedish.
We adapted the Emilia data processing pipeline and extracted ready-to-use TTS data from parliament speech recordings, resulting in 900 hours of Finnish and 5090 hours of Swedish TTS data. We choose parliament speech as the source because it is open-access and free to process and distribute as long as the original content is not modified. To further support the development of TTS for these languages, we propose unified evaluation sets following~\cite{chen2024f5, anastassiou2024seed}. Finally, we conduct benchmark experiments on two open-source TTS systems to demonstrate their applicability for model development and evaluation.

Our contributions can be summarized as follows:
\begin{itemize}
    \item We introduce \textit{Nord-Parl-TTS}, an open-source TTS dataset covering 900 hours of Finnish and 5090 hours of Swedish.
    \item We provide unified evaluation sets for Finnish and Swedish TTS.
    \item We benchmark two representative open-source TTS systems to demonstrate the applicability of the dataset and evaluation sets.
\end{itemize}
\noindent
The dataset, demos, and synthetic samples are available at: \\{\small\url{https://gryffindorli.github.io/nord-parl-tts-demo/}}.

\section{Related Work}
\label{sec:rel_work}

A brief summary of the existing Finnish and Swedish TTS datasets is illustrated in Table~\ref{tab:data_comp}. To our best knowledge, there is no existing public Swedish TTS dataset. Regarding Finnish,~\textit{Perso Synteesi}~\cite{perso} is a 20-hour speech dataset with 30 male and 30 female native speakers; \textit{CSS10}~\cite{park19c_interspeech} contains 10 hours of Finnish audiobooks read by a male speaker; and \textit{FinSyn}~\cite{finsyn} consists of approximately 60 hours of speech in various speaking styles by two female speakers. These datasets are on a small scale and are not suitable for training large speech generation models. Furthermore, \textit{Perso Synteesi}~\cite{perso} and \textit{FinSyn}~\cite{finsyn} are not conveniently available from open-sourced data hubs, like \textit{OpenSLR}\footnote{\url{https://www.openslr.org/index.html}} or \textit{HuggingFace datasets}\footnote{\url{https://huggingface.co/datasets}}.


\begin{table}[t]
\centering
\caption{Comparison of Nord-Parl-TTS with existing TTS datasets for Finnish and Swedish languages.}
\resizebox{\linewidth}{!}{
\begin{tabular}{lccc}
\toprule
\textbf{Language} & \textbf{Dataset} & \textbf{Data Source} & \textbf{Hours} \\ \hline
\multirow{4}{*}{\textbf{Finnish}} & Perso Synteesi~\cite{perso} & Studio Recording & 20 \\
& CSS10~\cite{park19c_interspeech} & Studio Recording & 10 \\
& FinSyn~\cite{finsyn} & Studio Recording & 60 \\
& \textbf{Nord-Parl-TTS} & \textbf{Parliament Recordings} & \textbf{900} \\
\midrule
\multirow{2}{*}{\textbf{Swedish}} & \multicolumn{3}{c}{No Public Dataset} \\
& \textbf{Nord-Parl-TTS} & \textbf{Parliament Recordings} & \textbf{5090}  \\ \hline
\end{tabular}
}
\label{tab:data_comp}
\end{table}

Finnish parliament recordings, which include nearly 3000 hours of parliament sessions between 2015 and 2020, have been transformed into the \textit{Finnish Parliament ASR corpus}~\cite{virkkunen2023_finparl}. However, further transforming the ASR corpus into a TTS dataset is not an optimal solution for several reasons. First, the transcripts are corrected for hesitations, repetitions, and slips of the tongue. They are also edited to replace spoken and spontaneous language with equivalent written language for clarity~\cite{virkkunen2023_finparl}. Second, the ASR corpus is segmented based on the word boundary, meaning that an utterance can start and end in the middle of a sentence. The corrected transcripts make it difficult for TTS models to learn the right pronunciations, and the word-level segmentation negatively affects the sentence-level prosody modeling. 

For Swedish, \textit{RixVox-v2}\footnote{\url{https://huggingface.co/datasets/KBLab/rixvox-v2}} provides a large-scale ASR dataset containing nearly 23000 hours of parliamentary speech and debate. The recordings are standardized to 16\,kHz mono audio, segmented into utterances of up to 30 seconds with sentence-level timestamps.

\section{Nord-Parl-TTS}
\label{sec:nord_tts}


This section elaborates on the details of the data processing pipeline and the approaches for curating evaluation datasets for Finnish and Swedish, respectively, as illustrated in Figure~\ref{fig:pipeline}.


\begin{figure*}[t]
    \centering
    \begin{subfigure}[b]{0.45\textwidth}
         \centering
         \includegraphics[width=\textwidth]{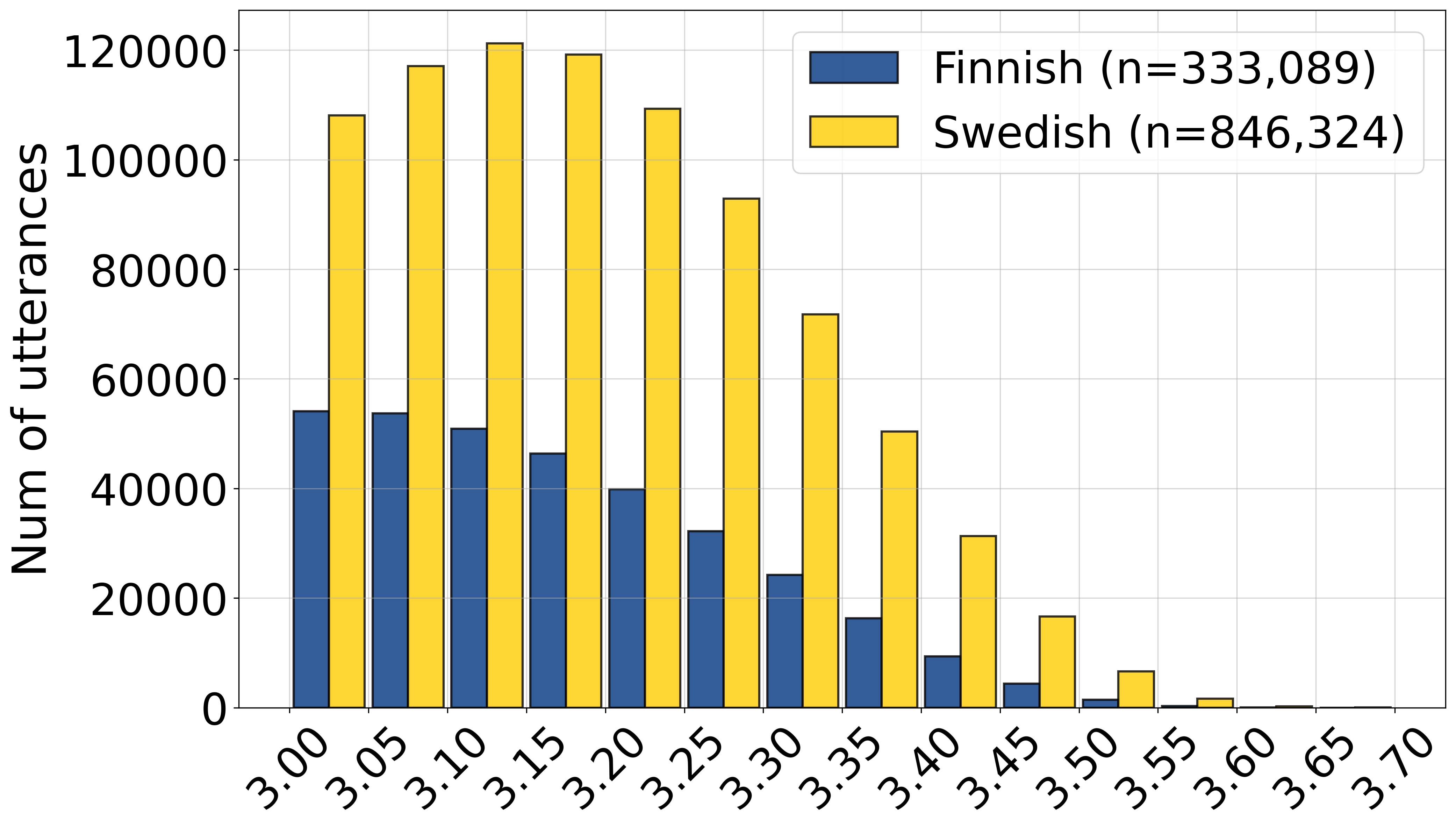}
         \caption{DNSMOS}
         \label{fig:mos}
    \end{subfigure}
    \hfill
    \begin{subfigure}[b]{0.45\textwidth}
         \centering
         \includegraphics[width=\textwidth]{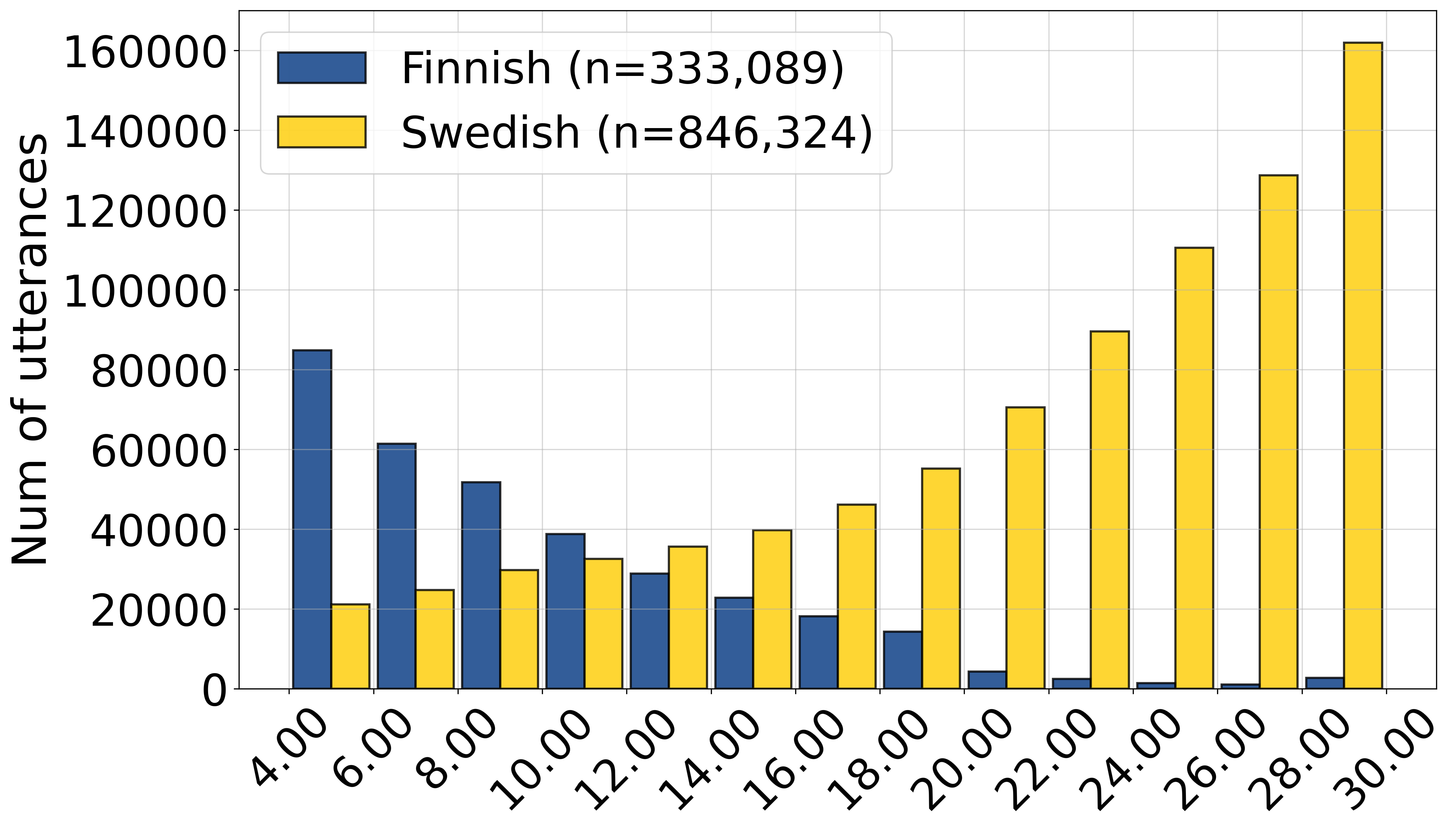}
         \caption{Duration}
         \label{fig:dur}
    \end{subfigure}
    \caption{The DNSMOS and duration distribution of the Nord-Parl-TTS dataset.}
    \label{fig:mos-dur}
\end{figure*}

\subsection{Finnish}
\label{subsec:fin}

Our Finnish processing pipeline mimics the Emilia pipeline~\cite{he2024emilia} with modifications, as shown by blue arrows in Figure~\ref{fig:pipeline}. We first extract the audio track from the video and standardize it to a 24\,kHz mono-channel format. It is then separated from noise using a pretrained \textit{UVR-MDX-Net Inst 3}\footnote{\url{https://github.com/TRvlvr/model_repo/releases/tag/all_public_uvr_models}} model and diarized using \textit{speaker-diarization-3.1} from Pyannote~\cite{Bredin23}. A fine-grained VAD is conducted on the diarized speech using \textit{Silero VAD}~\cite{SileroVAD}. Since the \textit{faster-whisper-large-v3}\footnote{\url{https://huggingface.co/Systran/faster-whisper-large-v3}} is not performant on Finnish, we decided to use an extra ASR model to validate the transcript. 
Specifically, we use a Wav2Vec2-large model pretrained and fine-tuned solely on Finnish~\cite{getman25_interspeech} as the second ASR model to validate the Whisper model's transcript based on the assumption that two different models would not make the same mistake. 
Therefore, if the WER between the normalized transcripts of these two models is less than $5\%$, we think the Whisper's transcript is confident and can be considered as a candidate for the final dataset.
We choose Whisper's transcripts here because both models hold a similar performance, while Whisper will restore capital letters and punctuations, which can be useful for TTS models. Following Emilia~\cite{he2024emilia}, the pipeline is finished with the DNSMOS~\cite{reddy2021dnsmos} prediction, with samples that have a DNSMOS P.835 OVRL score less than 3.0 to be excluded. The DNSMOS and duration distributions of the \textit{Nord-Parl-TTS} Finnish subset are shown in the blue pillars in Figure~\ref{fig:mos} and Figure~\ref{fig:dur}, respectively.


For the evaluation dataset, we curate 500 prompt-target pairs from \textit{Perso Synteesi}~\cite{perso}. We ensure a gender balance by sampling 250 prompts from male speakers and 250 prompts from female speakers. To avoid the error and model bias brought by the short sentences. All prompt and target speeches are between 3 seconds and 20 seconds, and the corresponding content has more than 10 characters. 

\subsection{Swedish}
\label{subsec:swe}

We process the \textit{RixVox-v2} for the Swedish dataset. The pipeline is shown in the yellow arrows in Figure~\ref{fig:pipeline}. We first resample the audio to 24\,kHz and use the same speech separation model in Section~\ref{subsec:fin} to clean out non-speech tracks. Then we cut the utterance using the existing sentence-level timestamp from the dataset if the gap between two sentences is longer than 2 seconds. The same speaker diarization model in Section~\ref{subsec:fin} is used to exclude segmented utterances with multiple speakers. To evaluate the intelligibility of the utterance, we transcribe the utterance using the Swedish Whisper-large~\cite{vesterbacka25_interspeech} and filter out those utterances that have a Word Error Rate (WER) larger than 10\%. 
We allow a WER of 10\% because the sentence-level transcripts and timestamps in \textit{RixVox-v2} contain similar corrections to those in the \textit{Finnish Parliament ASR corpus}~\cite{virkkunen2023_finparl}. The Swedish Whisper, trained on multiple large-scale Swedish datasets including \textit{RixVox-v2}~\cite{vesterbacka25_interspeech}, represents the content more faithfully.
Moreover, we wish to filter out unintelligible speech while keeping the scale of the dataset as large as possible. Finally, we filter out utterances with a DNSMOS P.835 OVRL score less than 3.0. This yields a 5090-hour Swedish dataset. The DNSMOS score distribution and the duration distribution of this dataset are shown in the yellow pillars in Figure~\ref{fig:mos} and Figure~\ref{fig:dur}, respectively. We notice that our pipeline for Swedish results in a monotonically increasing duration distribution in the dataset, which is different from existing datasets that are dominated by utterances in 5 to 15 seconds in duration.


We curate the Swedish evaluation dataset from CommonVoice~\cite{ardila2020common} Swedish 22.0. First, we keep utterances between 3 to 20 seconds with non-empty gender labels. Then we sample from each speaker a maximum of 30 utterances to create the pool of candidates while trying to balance the number of the utterances per gender and per speaker. After this, a native Swedish speaker listens to all these utterances using a bespoke curation tool and filters out utterances with unclear speech, low volume, strong non-native accents, and background noise. Then we form 500 prompt-target pairs as in Section~\ref{subsec:fin}.


\section{Experiments}
\label{sec:exp}


To validate the effectiveness of our \textit{Nord-Parl-TTS}, we train two non-autoregressive (NAR) TTS models, Matcha-TTS~\cite{mehta2024matcha} and F5-TTS~\cite{chen2024f5}, in the monolingual setting. Both models are diffusion-based. In particular, Matcha-TTS~\cite{mehta2024matcha} uses monotonic alignment search (MAS)~\cite{kim2020glow} to find the optimal alignment between input text and output mel-spectrogram, while F5-TTS~\cite{chen2024f5} learns implicit alignment through a stack of Diffusion Transformers (DiT)~\cite{peebles2023scalable} instead of explicit alignment modeling. No auto-regressive (AR) TTS models are included since we failed to find official open-source implementations for typical AR TTS models. For Matcha-TTS~\cite{mehta2024matcha}, we follow the implementation in~\cite{li25_ssw}, replacing the speaker embedding table with a pretrained \textit{SimAMResNet} Speaker Encoder from WeSpeaker~\cite{wang2024advancing} and strengthening the model with Classifier-free Guidance (CFG)~\cite{ho2021classifierfree}. We train the Swedish model with phoneme inputs using Phonemizer~\cite{Bernard2021} for Swedish G2P. Finnish on the other hand has a near one-to-one phoneme-to-grapheme mapping, so for the Finish model we use the character sequences directly. Both the Finnish and the Swedish models are trained on 1 Nvidia A100 GPU for 500k updates with a batch size of 64. 
For F5-TTS~\cite{chen2024f5}, we follow the \textit{F5TTS\_v1\_Base}\footnote{\url{https://github.com/SWivid/F5-TTS/blob/main/src/f5_tts/configs/F5TTS_v1_Base.yaml}} configuration. Both Finnish and Swedish models are trained using 24 AMD MI250X GPUs to maintain a global batch size of 308400 mel-spectrogram frames with 1.2M updates, as in the original paper.

We use the evaluation datasets described in Section~\ref{sec:nord_tts} to evaluate the performance of the TTS models objectively and subjectively. The CFG strength is set to $2.0$ for both models during inference time. For objective evaluation, we first synthesize the content of the target speech prompted by the prompt speech, then use ASR models to evaluate the Character Error Rate (CER) between the synthetic speech and the ground truth content. We also calculate the Cosine Speaker Similarity (SIM) between the synthetic speech and the prompt speech. We use Wav2Vec2-large~\cite{getman25_interspeech} for Finnish ASR and Swedish Whisper-large~\cite{vesterbacka25_interspeech} for Swedish ASR. Following~\cite{chen2024f5}, we use a WavLM-large-based~\cite{chen2022large} speaker verification model for SIM. All evaluations are conducted on a single Nvidia V100 GPU.

For subjective evaluations, we evaluate the human-likeness of the synthetic speech using Comparative Mean Opinion Score (CMOS) and speaker similarity using Speaker Mean Opinion Score (SMOS). We use CMOS instead of ``standard" Mean Opinion Score (MOS) to avoid perpetuating problematic practices~\cite{le2024limits}. For CMOS, we ask the participants whether sample A or sample B is more human-like, with a score between -3 to 3 and an increment of 1, where -3 means \textit{sample A is definitely more human-like}, 0 is \textit{unsure}, and 3 means \textit{sample B is definitely more human-like}. One of the two samples is the ground truth speech, and the other is the synthetic speech. For SMOS, we ask how sample B (synthetic speech) sounds like the same speaker as sample A (target speech), and the participants should rate between -2 and 2 with an increment of 1, where -2 represents \textit{definitely not the same speaker} and 2 represents \textit{definitely the same speaker}. We map SMOS back to a 1 to 5 scale when calculating the statistics. For both languages, we insert several language skill checks into each questionnaire, which instruct the participants to rate certain scores on certain questions in the testing language.
Our participants are recruited via \textit{Prolific}\footnote{\url{https://www.prolific.com/}}, a crowd-sourcing platform. For both languages, we screen the participants by their primary language, and with a history of at least 98\% acceptance rate. For Finnish, we also recruit students and employees from the university to ensure the desired number of participants. We exclude the submissions that fail the language skill checks from the final results.

\section{Results}
\label{sec:res}

The objective and subjective evaluation results are presented in Table~\ref{tab:res}. To exclude hallucinated synthetic utterances, we use CER $> 100\%$ as the threshold for filtering and calculate CER and SIM over the kept utterances. In particular, for F5-TTS-Base, 4 out of 500 utterances in the Finnish evaluation set and 18 out of 500 utterances in the Swedish evaluation set are hallucinated and filtered. For Matcha-TTS, only 1 out of 500 Swedish and 500 Finnish utterances is hallucinated and filtered.


We first discuss the objective evaluation results. In terms of CER, Matcha-TTS outforms F5-TTS-Base in both Finnish ($2.55\%$ vs $6.72\%$) and Swedish ($4.66\%$ vs $13.64\%$). Results in hallucination and CER indicate that the explicit alignment modeling reduces hallucination and improves the intelligibility of the synthetic speech, even though Matcha-TTS has fewer parameters, trained with a smaller batch size and fewer updates. For SIM, Matcha-TTS shows better speaker similarity than F5-TTS-Base ($0.566$ vs $0.538$) in Finnish, by directly conditioning on speaker embeddings. However, prompt-based F5-TTS-Base achieves better speaker similarity on Swedish, with a SIM of 0.53 over Matcha-TTS's SIM of 0.442.

For subjective evaluation, we invited 22 participants for Finnish and 20 participants for Swedish. Evaluation results show the models achieve acceptable performances in both languages. Specifically, for Finnish, Matcha-TTS achieves a CMOS of -1.93 and a SMOS of 2.49; F5-TTS-Base outputs Matcha-TTS for 0.5 on CMOS and 0.6 on SMOS. For Swedish, Matcha-TTS achieves a CMOS of -2.25 and a SMOS of 2.27, while F5-TTS-Base achieves a CMOS of -1.46 and a SMOS of 3.41. 
The evaluation results also demonstrate that an implicit alignment approach like F5-TTS~\cite{chen2024f5} improves the human-likeness at the cost of intelligibility.




\begin{table}[t]
    \centering
    \caption{Objective and subjective evaluation results of TTS systems. The best results of every column in each language are \textbf{bold}. The CMOS and SMOS scores are reported with 95\% Confidence Interval (CI).}
    \setlength\tabcolsep{4pt}
    \resizebox{\linewidth}{!}{
    \begin{tabular}{lccccc}
     \toprule
     \textbf{Language} & \textbf{Model} & \textbf{CER ($\downarrow$)} & \textbf{SIM ($\uparrow$)} & \textbf{CMOS ($\uparrow$)} & \textbf{SMOS ($\uparrow$)} \\
     \midrule
     \multirow{2}{*}{\textbf{Finnish}} & Matcha-TTS & \textbf{2.55\%} & \textbf{0.566} & -1.93 $\pm$ 0.12 & 2.49 $\pm$ 0.14 \\
     & F5-TTS-Base & 6.72\% & 0.538 & \textbf{-1.42 $\pm$ 0.14} & \textbf{3.12 $\pm$ 0.14} \\ 
     \midrule 
     \multirow{2}{*}{\textbf{Swedish}} & Matcha-TTS & \textbf{4.66\%} & 0.442 & -2.25 $\pm$ 0.11 & 2.27 $\pm$ 0.14 \\
     & F5-TTS-Base & 13.64\% & \textbf{0.530} & \textbf{-1.46 $\pm$ 0.13} & \textbf{3.41 $\pm$ 0.15} \\ 
    \bottomrule
    \end{tabular}
    }
    \label{tab:res}
\end{table}



\section{Conclusions}
\label{sec:conclu}


In this paper, we present \textit{Nord-Parl-TTS}, an in-the-wild TTS dataset for Finnish and Swedish derived from parliament speech recordings. Open-sourced TTS models trained on this dataset achieve acceptable synthesis results and proves our dataset is applicable for TTS training. 
In addition to refining the tooling and making curation tools and data visualization available, our future work includes extending the dataset to other Nordic languages like Danish and Norwegian, providing more benchmark models, and further improving dataset quality. 


\section{Acknowledgement}

This work was a part of the Ministry of Education and Culture’s Doctoral Education Pilot in Finland, under Decision No. VN/3137/2024-OKM-6 (The Finnish Doctoral Program Network in Artificial Intelligence, AI-DOC). We acknowledge the EuroHPC Joint Undertaking for awarding this project access to the EuroHPC supercomputer LUMI, hosted by CSC (Finland) and the LUMI consortium through a EuroHPC Regular Access call. We acknowledge the computational resources provided by the Aalto Science-IT project.

\bibliographystyle{IEEEbib}
\ninept
\bibliography{strings,refs}

\end{document}